\documentclass{ws-procs10x7}
\def\simge{\mathrel{%
   \rlap{\raise 0.511ex \hbox{$>$}}{\lower 0.511ex \hbox{$\sim$}}}}
\def\simle{\mathrel{
   \rlap{\raise 0.511ex \hbox{$<$}}{\lower 0.511ex \hbox{$\sim$}}}}

\begin{document}

\title{Color Glass Condensate in QCD at High Energy}

\author{Kazunori\ \ Itakura}

\address{Service de Physique Th\'eorique, CEA/Saclay, F-91191, Gif-sur-Yvette, France\\
E-mail: itakura@spht.saclay.cea.fr }

\twocolumn[\maketitle\abstract{I give a brief review 
about the color glass condensate, which is the universal form of hadrons 
and nuclei at high energies.}]

\section{The Color Glass Condensate}\vspace*{-3.3mm}
The main purpose of this talk is to convince you that 
{\it high energy limit of QCD is the Color Glass Condensate}.\cite{review}
More precisely, when the energy of hadronic scattering is very large, 
the participating hadrons or nuclei behave not as  usual states made of 
valence particles, but as new form of matter called the color 
glass condensate (CGC).
This name is after the following observations.
First of all, it is made of gluons ("small-$x$ gluons")
which have {\bf color} and carry small fractions $x\ll 1$ of the 
total momentum.
Next, these small-$x$ gluons are created by slowly moving color 
sources (partons with larger $x$) which are distributed randomly 
on the  two dimensional disk (the Lorentz contracted hadron). 
This is like a {\bf glass} whose constituents are disordered and 
appear to be frozen in short time scales. 
Lastly, the density of small-$x$ gluons becomes very large until 
it is saturated to some value. Typically the occupation number of gluons 
is of ${\cal O}(1/\alpha_s) \gg 1$ at saturation, 
which is like a {\bf condensate} of bosons. 
As the scattering energy is increased, the hadrons undergo
multiple production of small-$x$ gluons, and eventually become
the CGC. 

Note that the claim above is as correct as the statement
about the other limits of QCD: high 
temperature/density limit of QCD is the QGP/color superconductor,
which now everyone believes true. In the same sense, if one goes 
to high energy limit in QCD, one will necessarily encounter the CGC. 
Note also that these three different limits allow for
weak-coupling descriptions powered by sophisticated 
resummation schemes.

\section{Gluon saturation and unitarity}\vspace*{-3.3mm}
Let us explain how the CGC appears with increasing energy.
When the gluon density is not so high, change of the 
gluon distribution with increasing energy (or decreasing $x$) 
is described by the BFKL evolution 
equation:\cite{BFKL}
\begin{equation}\hspace*{-2mm}
\frac{\partial}{\partial \tau}{\cal N}_\tau(k) 
= \bar\alpha_s K_{\rm BFKL}\otimes {\cal N}_\tau(k),
\end{equation}
where $\bar\alpha_s={\alpha_sN_c}/{\pi}$, 
${\cal N}_\tau(k)$ is the dipole-proton scattering amplitude
(or unintegrated gluon distribution of the target proton),
$\tau=\ln 1/x$ is the rapidity, and 
$\bar\alpha_s K_{\rm BFKL}$ is the kernel 
representing the probability of splitting of one dipole into two.
This is essentially a linear differential equation, and its 
solution at asymptotically large energies shows exponential 
growth of ${\cal N}_\tau(k)$. Namely, multiple production of 
gluons occurs endlessly. This, however,
violates the unitarity bound for the cross section (or, ${\cal N}_\tau\le 1$) 
and the BFKL equation must be modified so as not to violate the 
unitarity. In fact, what is missing in the BFKL equation 
is the {\it recombination process} of two gluons into one, which cannot be ignored 
when the gluon density is high. Note that this process 
effectively reduces the speed of growth. 
Once this is included, the BFKL equation is replaced by 
the Balitsky-Kovchegov (BK) equation:\cite{BK} 
\begin{equation}\hspace*{-2mm}
\frac{\partial}{\partial \tau}{\cal N}_\tau(k)=
\bar\alpha_s K_{\rm BFKL}\otimes \left(
{\cal N}_\tau(k)-{\cal N}^2_\tau(k)\right)\!.\!\!
\end{equation}
Compared to the BFKL solution, 
the solution to this equation is drastically changed due to the presence 
of the nonlinear term. Indeed, 
it exhibits saturation (unitarization) of the amplitude and a kind of 
universality, as I will explain below in a simple example.

\vspace*{-0.3cm}
\subsection{Analogy with population growth}\vspace*{-0.2cm}
In order to understand what happens in the BK equation, 
let us ignore the transverse dynamics for the time being.
This simplification allows us to find an interesting analogy 
with the problem of population growth. Long time ago, Malthus discussed  
that growth rate of population should be proportional to 
the population itself, and proposed a simple linear equation 
for the population density $N(t)$:\vspace*{-0.1cm}
\begin{equation}\vspace*{-0.1cm}
\frac{d}{dt}N(t)=\alpha N(t).
\end{equation}
Its solution $N(t)\!=\!N_0\, {\rm e}^{\alpha t}$ shows exponential 
growth known as the "population explosion."
However, as the number of people increases, this 
equation fails to describe the actual growth. 
This is because various effects such as lack of foods reduce the 
speed of growth. One can effectively represents such effects by 
replacing the growth constant $\alpha$ by $\alpha(1-N)$ which 
decreases with increasing $N$. This yields the famous 
{\it logistic equation} first proposed by Verhulst:\vspace*{-0.1cm}
\begin{equation}\vspace*{-0.1cm}
\frac{d}{dt}N(t)=\alpha \left(N(t)-N^2(t)\right).
\end{equation}
This {\it nonlinear} equation can be solved analytically.
The similarity of these equations to our problem is rather 
trivial: $N$ and $t$ correspond to the scattering amplitude and the 
rapidity, respectively.
In Fig.~1, we show the solutions to eq.~(4)
with different initial conditions at $t=0$, together with 
the corresponding solutions to the linear equation (3).
We can learn much from this result. First of all, at early time 
$t\sim 0$,
the solution shows exponential growth as in the linear case.
However, as $N(t)$ grows, the nonlinear term ($\sim N^2$)
becomes equally important, and the speed of growth is 
reduced. Eventually at late time, the solution approaches to 
a constant which is determined by the asymptotic condition 
$dN/dt=0$. 
This corresponds to the {\it saturation} and {\it unitarization} 
of the gluon number.
Next, note that two solutions of the logistic equation 
with different initial conditions approach to each other, and converge 
to the same value, while deviation of two solutions of 
the linear equation expands as time goes. Namely, in the 
logistic equation, the initial 
condition dependence disappears as $t\to \infty$, which is 
related to the {\it universality} of the BK equation.
This analogy may look oversimplified, but actually most of 
the phenomena seen in the BK equation can be identified in this 
simple example, as far as energy dependence of the solution is 
concerned.\footnote{Indeed, the logistic equation (4) is obtained 
in the homogeneous 
approximation of the FKPP equation, which is essentially equivalent 
to the BK equation.\cite{Munier}}
\vspace*{-0.4cm}

\begin{figure}
\epsfxsize160pt
\figurebox{120pt}{160pt}{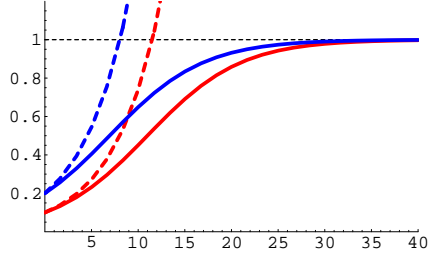}\vspace*{-0.3cm}
\caption{Solutions to eq.~(4) [solid lines] and eq.~(3) [dashed lines], 
with different initial conditions.}
\label{fig1}\vspace*{-0.5cm}
\end{figure}

\section{Saturation scale}\vspace*{-0.3cm}
Let us turn to the transverse dynamics which was ignored 
above. The most important quantity associated with 
the transverse dynamics is the {\it saturation scale} $Q_s$, 
which gives the border line between the saturated (nonlinear) 
and non-saturated (linear) regimes. Physically, it corresponds to 
(inverse of) the typical 
transverse size of gluons when the transverse plane of a hadron 
or nucleus is filled with gluons. 
More precisely, it is given by the scale $Q^2$ satisfying
\vspace*{-0.15cm}
\begin{equation}\hspace*{-2mm}\vspace*{-0.15cm}
\sigma\cdot \rho =1,\  {\rm with} \ 
\sigma=\frac{\alpha_s}{Q^2}, \ \rho= \frac{xG(x,Q^2)}{\pi R^2}.
\end{equation}
Here $\sigma$ is the cross section between the external 
probe and a gluon, and $\rho$ is the gluon density
per unit transverse area of the target with its radius $R$. 
Note that the saturation scale depends upon both 
$x$ and atomic number $A$. Its $x$ and $A$ dependences can be 
precisely determined by the BK equation or the linear BFKL equation 
with saturation boundary condition. When $x$ is not too small, 
one finds\cite{GLR,IIM,MT,Dionysis} \vspace*{-0.25cm}
\begin{equation}\vspace*{-0.15cm}
Q_s^2(x,A)\propto A^{1/3} x^{-\lambda}
\end{equation}
where $\lambda$ is given by $4.88 \bar\alpha_s$ from the LO-BFKL, 
and $\lambda\sim 0.3$ for $x=10^{-2}-10^{-4}$ from (resummed) NLO-BFKL.
Also, for the running coupling case, $A$ dependence of $Q_s$ disappears 
at very large energy (small $x$).\cite{Mueller}
This particular dependence upon $x$ and $A$ leads to an interesting 
observation that the saturation scales for the deep inelastic 
scattering (DIS) at HERA and for the Au-Au collisions at RHIC are of 
the same order. 
Therefore, if one finds saturation effects in the HERA data, then there 
is enough reason to expect similar things in the RHIC data.
\vspace*{-0.4cm}
\section{Geometric scaling}
\vspace*{-0.3cm}
Geometric scaling\cite{GS} is one of the most significant 
experimental facts which provides an indirect evidence of the CGC. 
This is a new scaling phenomenon at small $x$
meaning that the total $\gamma^*$-proton cross section 
$\sigma_{\rm total}^{\gamma^* p}$ in DIS depends upon 
$Q^2$ and $x$  only via their specific combination 
$\xi\equiv Q^2 R_0^2(x)$, with $R_0^2(x) \propto x^{\lambda},\ 
\lambda\sim 0.3$. 
Recall that $1/Q$ is the transverse size of gluons measured 
by the virtual photon, and, particularly, $1/Q_s$ is 
the gluon size when the transverse plane of the proton is 
filled with gluons. Therefore, if we identify $R_0(x)\equiv 1/Q_s(x)$,
then the scaling variable $\xi^{-1}$ turns out to be 
a ratio between the transverse areas of these gluons 
$\xi^{-1}=\sigma_{g}/\sigma_{sat}$. 
Moreover, if gluons are distributed homogeneously over the transverse plane,
which is realized in the saturated regime, $\xi^{-1}$ corresponds to the 
{\it number of covering times}. For example, when $\xi=1$, the 
transverse plane is covered by gluons {\it once} which, however, 
can be realized by various size of gluons. Since what matters in 
the saturated regime is the effective number of overlapping, 
it is natural that different kinematics can give the same value 
of cross section if the number of covering is the same. 
This is nothing but the geometric scaling!
Furthermore, it is remarkable that the theoretical calculation for 
the $x$ dependence of $Q_s(x)$ is consistent with the experimentally
determined one $Q_s^2(x)\propto x^{-0.3}$. 

The scenario discussed so far holds naturally in the saturation 
regime. On the other hand, the data show the geometric scaling 
up to higher values of $Q^2$ ($\sim$ 100 GeV$^2$), which is the 
non-saturated regime described by the linear BFKL equation.
Indeed, this is again well understood in the context of CGC.
The geometric scaling appears if one solves the BFKL equation 
{\it with a saturation boundary condition}. 
As we depart from the saturation line toward the linear regime, 
the effect of saturation becomes weaker and weaker and eventually 
disappears. In fact, one can determine the window for
the geometric scaling:\cite{IIM}\vspace*{-1mm}
\begin{equation}\vspace*{-1mm}
Q_s^2(x)\ \simle \ Q^2\  \simle\ Q_s^4(x)/\Lambda^2_{\rm QCD}.
\end{equation}
Hence we recognized that there is a qualitatively different regime 
in between the CGC and DGLAP regimes, as is shown in Fig.~2.
\vspace*{-0.3cm}
\begin{figure}
\epsfxsize200pt  \vspace*{-0.4cm}
\figurebox{120pt}{160pt}{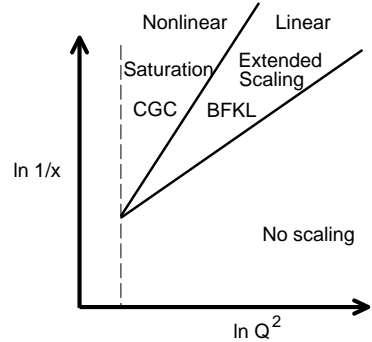}\vspace*{-1.1cm}
\caption{"Phase diagram" of a proton in DIS.}
\label{fig2}\vspace*{-0.5cm}
\end{figure}

\section{The CGC confronts experiments}
\vspace*{-0.3cm}
Our understanding of the CGC has been developed with the experimental 
results at HERA and RHIC, as we already saw above. 
Below I briefly explain some of the attempts to 
describe/understand the other experimental data from the viewpoint of 
CGC.
\vspace*{-0.3cm}
\subsection{HERA small $x$ data}\vspace*{-0.2cm}
DIS at small $x$ is the cleanest process for measuring 
the saturation effects in gluon distribution of the proton.
Starting from the pioneering work by Golec-Biernat and 
W\"usthoff,\cite{GBW} there are several efforts to describe the 
HERA DIS data such as the $F_2$ structure function at small $x$,
in the context of gluon saturation. 
Here let us explain the "CGC fit"\cite{CGC_fit} which is one of the most 
successful fits of the small $x$ data based on QCD.
The CGC fit is constructed so as to contain two approximate 
solutions to the BK equation, which are valid 
in the saturation and BFKL regimes, respectively.
In particular, the solution in the linear BFKL regime 
shows the geometric scaling and its (small) violation. 
With only three parameters, the CGC fit provides a very nice 
fit for the $F_2$ structure function with $x<10^{-2}$ and 
in $0.045 < Q^2 < 45\,$GeV$^2$. 
Meanwhile, it turned out that
this fit works reasonably well even for other observables such as 
$F_2^{\rm diff}$ and the vector meson production.
\vspace*{-0.3cm}
\subsection{RHIC Au-Au data}
\vspace*{-0.2cm}
The CGC provides the initial condition for the heavy ion collision. 
Information of the initial state could  still be seen in the 
final observed data. It should be noticed that most of the produced 
particles have small momenta less than 1 GeV which is of the same order 
as $Q_s$ in RHIC. This observation suggests that 
effects of saturation may be visible in bulk quantities such as 
the multiplicity. Indeed, the CGC results\cite{Dima_Levin} 
 for the pseudo-rapidity and centrality dependences of the multiplicity
are in good agreement with the data.
\vspace*{-0.3cm}
\subsection{RHIC d-Au data}
\vspace*{-0.2cm}
The CGC has recently come under the spotlight with the experimental 
results\cite{Brahms} for the 
nuclear modification factor in the deuteron-Au collisions 
measured by the Brahms experiment at RHIC. The data show 
enhancement of the ratio at mid-rapidity (the Cronin effect) 
and suppression at forward rapidities. In fact, the global behavior of the 
data is qualitatively consistent with the predictions made by 
the CGC.\cite{Dima,Armesto}
More recently, detailed analysis of the ratio was done\cite{IIT} 
and it has been clarified that the Cronin effect is due to the 
multiple collision and re-distribution of the gluons 
which is properly described by the McLerran-Venugopalan model, 
and that the high $p_\perp$ suppression 
is induced by the mismatch of the evolution speed between 
the proton (deuteron) and the nucleus. The nucleus is closer to 
saturation and evolves slower than the proton.


\vspace*{-0.4cm}
\begin{thebibliography}{99}
\vspace*{-0.25cm}
\bibitem{review}
For a recent review, see E.~Iancu and R.~Venugopalan, hep-ph/0303204.

\bibitem{BFKL}L.\,Lipatov, {\it Sov.~J.~Nucl.~Phys.}~{\bf 23}, 338 
(1976),\,E.\,Kuraev,~L.\,Lipatov~and~V.\,Fadin, {\it Sov. Phys. JETP}
 {\bf 45},\,199\,(1977), I.~Balitsky and L.\,Lipatov, 
{\it Sov.~J.~Nucl.~Phys.} {\bf 28}, 822 (1978).
\bibitem{BK}I.\,Balitsky, {\it Nucl.~Phys.}~{\bf B463},~99~(1996),
Y.\,Kovchegov, {\it Phys. Rev.} {\bf D60}, 034008 (1999).
\bibitem{Munier}S.~Munier and R.~Peschanski, {\it Phys. Rev. Lett.}
{\bf 91}, 232001 (2003). 
\bibitem{GLR}L.~Gribov, G.~Levin, and M.~Ryskin, 
{\it Phys. Rept.} {\bf 100}, 1 (1983). 
\bibitem{IIM}E.~Iancu, K.~Itakura, and L.~McLerran, {\it Nucl. Phys.} 
{\bf A708}, 327 (2002).
\bibitem{MT}A.~Mueller and D.~Triantafyllopoulos,
{\it Nucl. Phys.} {\bf B640}, 331 (2002) 331.
\bibitem{Dionysis}D.Triantafyllopoulos,~{\it Nucl.~Phys.}~{\bf B648},
 293 (2003).
\bibitem{Mueller}\hspace*{-1mm}A.Mueller,~{\it Nucl.~Phys.}~{\bf A724},~223~(2003)


\bibitem{GS}A.~Stasto, K.~Golec-Biernat and J. Kwie-cinski,
{\it Phys. Rev. Lett.}  {\bf 86}, 596 (2001).

\bibitem{Dima_Levin}
D.~Kharzeev and E.~Levin, {\it Phys. Lett.} {\bf B523}, 79 (2001).


\bibitem{GBW}K.~Golec-Biernat and M.~W\"usthoff,
{\it Phys. Rev.} {\bf D59}, 014017 (1999), {\it ibid.} 
{\bf D60}, 114023 (1999).
\bibitem{CGC_fit}
E.~Iancu, K.~Itakura and S.~Munier,
{\it Phys. Lett.}  {\bf B590}, 199 (2004).




\bibitem{Brahms}I.~Arsene {\it et al.}  [BRAHMS Collaboration],
arXiv:nucl-ex/0403005.
\bibitem{Dima}D.~Kharzeev,~Y.~Kovchegov~\&~K.~Tuchin
{\it Phys. Rev.} {\bf D68}, 094013 (2003).
\bibitem{Armesto}
J.~Albacete, 
N.~Armesto, A.~Kovner, C.~Salgado and U.~Wiedemann,
{\it Phys. Rev. Lett.}  {\bf 92}, 082001 (2004)
\bibitem{IIT}
E.~Iancu, K.~Itakura and D.~Triantafyllopoulos,
{\it Nucl.~Phys.}~{\bf A742},~182~(2004).

\end{thebibliography}
\end{document}